\documentclass{article}
\usepackage{hiph-art}
\def\rightmark{The role of noise and dissipation 
in the hadronization of the QGP}
\def\leftmark{E. S. Fraga}
 
\title{The role of noise and dissipation 
in the hadronization of the quark-gluon plasma} 
\authors{ 
{Eduardo S. Fraga$^1$ %
\index{Fraga, E. S.} % Abbreviated names of the author(s),
}\\[2.812mm]
{\normalsize
\hspace*{-8pt}$^1$ Instituto de F\'\i sica, 
Universidade Federal do Rio de Janeiro\\ 
C.P. 68528, Rio de Janeiro, RJ 21941-972, Brazil\\[0.2ex] 
}}
 
\abstract{We discuss the role of noise and dissipation in the 
explosive spinodal decomposition scenario of hadron production 
during the chiral transition after a high-energy heavy ion 
collision. We use a Langevin description inspired by 
nonequilibrium field theory to perform real-time lattice 
simulations of the behavior of the chiral fields. Preliminary 
results for the interplay between additive and multiplicative 
noise terms, as well as for non-Markovian corrections, are also 
presented. }

 
\makeindex
\begin{document}
 
\maketitle
%\setcounter{page}{1}

%%%%%%%%%%%%%%%%%%%%%%%%%%%%%%%%%%%%%%%%%%%
\section{Introduction}
\label{intro}

For high enough values of temperature, strongly interacting 
matter should be in a quark phase due to asymptotic freedom. In fact, 
finite-temperature lattice QCD simulations provide strong evidence of a 
deconfined quark-gluon plasma (QGP) phase at sufficiently high 
temperature \cite{Laermann:2003cv}. Moreover, relativistic high-energy 
heavy ion collisions can probe strongly interacting matter 
under extreme conditions. They provide valuable information on the new 
state of matter that seems to have been created according to 
recent data from experiments at BNL-RHIC \cite{QM2004}, even if 
its true nature is still uncertain.

As the QGP presumably created in a heavy-ion collision expands, 
it cools down and, at $T_c\sim 150~$MeV 
according to the Lattice \cite{Laermann:2003cv}, undergoes a 
phase transition (or a crossover) back to hadronic matter. 
Results from CERN-SPS and BNL-RHIC feature what has been called
\textit{sudden hadronization} \cite{sudden} or \textit{explosive
behavior} \cite{explosive} in the hadronization process of the
expanding QGP and seem to favor a fast (explosive) spinodal
decomposition scenario as the mechanism of phase conversion. 
This should be confronted to the process of bubble nucleation, whose 
time scales are larger due to the existence of a barrier to 
overcome \cite{review}. Possible signatures of 
the explosive behavior in high-energy nuclear collisions were 
proposed in Ref. \cite{randrup}. 

Effective field theory models for the chiral and the deconfinement 
transitions in QCD also indicate that most of the plasma is quenched 
into the spinodal region, due to the fast expansion of the system as 
compared to the low nucleation rate, then undergoes an explosive 
phase conversion 
\cite{explosive,polyakov,Csernai:1995zn,Scavenius:1999zc,Scavenius:2001bb,Paech:2003fe}. 
This leads to what we will refer to as the explosive spinodal 
decomposition scenario. Although an expansion term in the evolution 
equation for the order parameter plays a role somewhat similar to 
dissipation, most studies do not take into account genuine dissipative 
effects due to the interactions with the medium. 
However, dissipation effects have proved to be important, for 
instance, in the context of disoriented chiral condensate (DCC) 
formation in heavy ion collisions 
\cite{biro,Greiner:1996dx,dirk,krishna,agnes,dcc} 
and could be an important ingredient in the explosive spinodal 
decomposition scenario.

Recently, we have considered the effects of dissipation on 
hadron production during the QCD transition after 
a high-energy heavy ion collision in the simplest fashion \cite{FK1}. 
Using a phenomenological Langevin description for the time evolution 
of the order parameter, inspired by microscopic nonequilibrium field 
theory results \cite{Greiner:1996dx,dirk,Gleiser:1993ea,Berera:1998gx}, 
we performed real-time lattice simulations for the behavior 
of the inhomogeneous chiral field. We concluded that the effects of
dissipation could be dramatic even for very conservative
assumptions. It was shown that even if the system quickly 
reaches this unstable region there is still no guarantee that it 
will explode. 

Nevertheless, the description implemented in Ref. \cite{FK1} captures 
only a small part of the much richer spectrum of possibilities for 
dissipation and noise effects in the dynamics of phase transitions. 
Part of this richness is captured by a linear response nonequilibrium 
quantum field theory analysis, which usually provides not only 
different contributions to noise and dissipation terms, 
but also complicated memory kernels instead of simple Markovian terms 
proportional to the time derivative of the field 
\cite{Greiner:1996dx,dirk,Gleiser:1993ea,Berera:1998gx}. 

In what follows, we present our findings from a traditional Langevin 
analysis of the evolution of the order parameter for the chiral 
transition, and discuss preliminary results including multiplicative 
noise \cite{FKR} and memory effects \cite{FKKP}. A robust feature that 
comes about from this study is that all the contributions tend to add up 
to delay the process of thermalization.

%%%%%%%%%%%%%%%%%%%%%%%%%%%%%%%%%%%%%%%%%%%%
\section{Dissipation versus explosive decomposition}
\label{simple-langevin}  

We consider the real-time dynamics of chiral
symmetry breakdown of a QGP created in a high-energy heavy ion
collision \cite{FK1}. We assume the system to be characterized by a
coarse-grained free energy
\begin{equation}
F(\phi,T)=\int d^3x \left[ \frac{1}{2}(\nabla\phi)^2 +
V_{eff}(\phi,T)  \right] \; , \label{free-energy}
\end{equation}
where $V_{eff}(\phi,T)$ is an effective potential of the
Landau-Ginzburg form whose coefficients depend on the temperature,
and $\phi(\vec{x},t)$ is a real scalar field which plays the role
of an order parameter that is {\em not} conserved, such as the
chiral condensate. To model the mechanism of chiral symmetry
breaking found in QCD, we adopt the linear $\sigma$-model coupled
to quarks, whose standard Lagrangian can be found, for instance,
in Ref.~\cite{Scavenius:2001bb}. This approach is widely used in 
the literature and its specificities imply no major limitations 
to our main results. Quarks are integrated out yielding the
effective potential $V_{eff}(\phi,T)$. The pion directions play no 
major role in the process of phase conversion we are considering, so 
we focus on the sigma direction represented by the field 
$\phi$ \cite{Scavenius:2001bb}. For simplicity, we ignore effects 
due to the finite size of the plasma \cite{Fraga:2003mu}.

The framework for the dynamics is assumed to be given by the following
Langevin equation:
\begin{equation}
\frac{\partial^2\,\phi}{\partial t^2} - 
\nabla^2 \phi + \eta \frac{\partial\phi}{\partial t} +
\frac{d V_{\rm eff}(\phi)}{d\phi}=\xi(\vec{x},t) \, ,
\label{langevin-simple}
\end{equation}
where $\phi$ is a real scalar field and $\eta$, which can be seen 
as a response coefficient that defines time scales for the system 
and encodes the intensity of dissipation, is usually taken to be a 
function of temperature only, $\eta=\eta (T)$. The function 
$\xi(\vec{x},t)$ represents a stochastic (noise) force, assumed 
Gaussian and white, so that
$\langle \xi(\vec{x},t) \rangle = 0$ and
$\langle \xi(\vec{x},t)\xi(\vec{x'},t') \rangle =
2\,\eta T \delta(\vec{x}-\vec{x'})\delta(t-t')$,
according to the fluctuation-dissipation theorem.

In our numerical simulations we solve Eq.~(\ref{langevin-simple}) 
on a cubic space-like lattice with $64^3$ sites under periodic boundary
conditions, with a lattice spacing of $a~=~0.91$~fm. We use a
semi-implicit finite-difference scheme for the time evolution and
a Fast Fourier Transform for the spatial dependence \cite{mimc}.
Temperature is fixed to the spinodal value $T_{sp}\approx
108~$MeV \cite{Scavenius:2001bb}. 
We perform several runs starting from different random
initial configurations around the inflexion point of $V_{eff}$
which happens at $\phi_0 \approx 0.162\,T$ and then average the
results from the different initial configurations. For time steps
of $\Delta t~=~0.001/T$ the results become independent of the
lattice spacing once it is smaller than $a~\simeq~1$~fm.

We show results of simulations for three different values of the
dissipation coefficient, namely $\eta/T = 0$, $2$ and $4$.
It can be argued that the response coefficient has the form
$\eta(T) \approx 2\,T/b$, where $b$ is a number of order one to
first approximation \cite{Kajantie:1992uk}. The cases considered
provide a conservative band around the value $\eta(T) \approx
2\,T$ to illustrate the effect of dissipation.

\begin{figure}[htb]
\vspace*{-0.1cm}
                 \insertplot{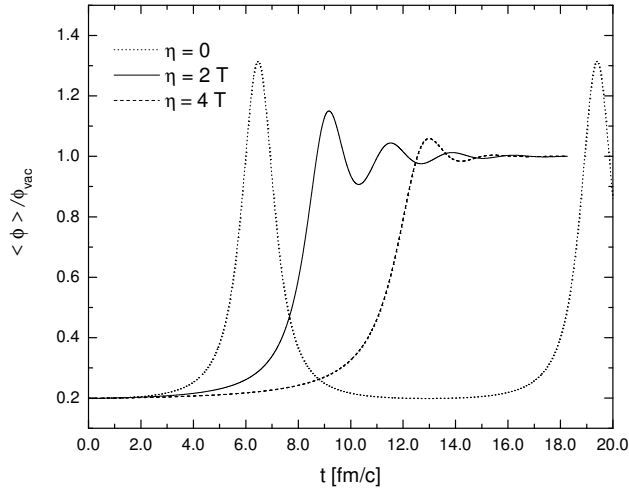}
\vspace*{-0.1cm}
\caption[]{Average value of the chiral field $\phi$ in units of its
vacuum value $\phi_{vac}$ as a function of time for $\eta/T= 0,
2, 4$.}
\label{additive}
\end{figure}

In Figure~\ref{additive} we show the average value of $\phi$ in
units of its vacuum value, $\phi_{vac}$, as a function of time for
the three different values of $\eta$ mentioned above. 
The results clearly show that even for a very conservative value
of dissipation, $\eta = 2\,T$, the effect can be dramatic. For
this value of $\eta$, dissipation retards the time evolution of
$\phi$ towards its vacuum value in $\sim 100\%$ compared to the
case with $\eta=0$. The important point to be noted here is that
for expansion times of the order of $5~$fm/c, which is of the
order of the time scales for RHIC collisions, there might be
not enough time for the onset of the spinodal
explosion.

%%%%%%%%%%%%%%%%%%%%%%%%%%%%%%%%%%%%%%%%%%%%%%
\section{Improved Langevin approach}
\label{improved}

Although phenomenological equations of the form of Eq.
(\ref{langevin-simple}) are widely used in the literature, formal
derivations of effective equations of motion based on a more
complete field-theoretic description of nonequilibrium  dissipative
dynamics \cite{dirk,Gleiser:1993ea,Berera:1998gx} show that much more
complicated equations emerge. In general they are non-local
equations of motion, with colored, non-Markovian dissipative
and noise kernels, exhibiting both noise and dissipation terms
that depend on the field amplitude. It is then important to
analyze the possible outcome of dynamics emerging from the more
realistic equations as compared to the case of the simpler
phenomenological equation (\ref{langevin-simple}). 
In fact, on very general physical
grounds, one expects that dissipation effects should depend on the
local density $\sim\phi^2\dot\phi$ and, accordingly, the noise
term should contain a multiplicative piece $\sim\phi$. These
results emerge as approximations to the
effective equations of motion derived {\it e.g.} in Refs.
\cite{dirk,Gleiser:1993ea,Berera:1998gx}. 

Motivated by the results referred above, 
we consider an improved Langevin framework which 
includes the effects of multiplicative noise and density-dependent 
dissipation terms in the description of the time evolution of the 
order parameter. To assess the effect of the new contributions on 
the relevant time scales for phase ordering, one can adopt a 
scalar $\lambda\phi^4$ theory in the broken phase, whose standard 
Lagrangian density has the form:
\begin{equation}
\mathcal{L} = \frac{1}{2}(\partial_{\mu}\phi) (\partial^{\mu}\phi)+
\frac{1}{2}m^2\phi^2 - \lambda\phi^4\;,
\label{lagrangian}
\end{equation}
where $\phi$ is a scalar field playing the role of the order parameter
and $\lambda > 0$. We perform $(3+1)$-dimensional real-time
lattice simulations to study the behavior of the inhomogeneous
scalar field, taking into account lattice counterterms that guarantee
lattice-size independence \cite{krishna,Borrill:1996uq,Gagne:1999nh}.

To define an initial value problem for the evolution of the 
order parameter, one can assume that the system was quenched
from a high-temperature environment in which the sign of the mass term
was the opposite, corresponding to a symmetric phase, down to a 
temperature below some critical temperature of symmetry restoration
($T < T_c$), in which the effective potential has the form shown in
the Lagrangian density \ref{lagrangian}.

In our analysis, the time evolution of the field $\phi(\vec{x},t)$ at
each point in space and its approach to equilibrium will be dictated
by an improved Langevin equation of the form
\begin{equation}
\left(\frac{\partial^2}{\partial t^2} - \nabla^2\right) \phi(\vec{x},t) +
[\eta_1 +  \eta_2\phi^2(\vec{x},t) ] \frac{\partial\phi(\vec{x},t)}{\partial t} +
V'(\phi)=\xi_1(\vec{x},t) + \phi(\vec{x},t) \xi_2(\vec{x},t) \; ,
\label{langevin}
\end{equation}
which models the local approximation for the effective equation of motion
derived e.g. in Ref. \cite{Gleiser:1993ea}. In Eq. (\ref{langevin}),
$\eta_1$ and $\eta_2$ will be taken to be functions of
temperature only, $\eta_i=\eta_i (T)$. The functions
$\xi_1(\vec{x},t)$ and $\xi_2(\vec{x},t)$ represent stochastic 
(noise) forces, assumed Gaussian and white.

Eq. (\ref{langevin}) could, in principle, be obtained from a
microscopic field-theoretic description of the real-time
nonequilibrium dynamics of the chiral field at finite 
temperature \cite{dirk,Gleiser:1993ea}.
The noise and dissipation terms, which originate from quantum
fluctuations, are engendered by either self-interactions of the
chiral field or coupling to one or more different fields that play
the role of a heat bath, provided one incorporates higher-order
terms in the computation of the effective equation of motion for
$\phi(\vec{x},t)$. In fact, it is well-known that one has to go up
to two-loop corrections in order to pick up imaginary parts in the
self-energy associated with viscosity and dissipation
\cite{dirk,agnes,Gleiser:1993ea,Berera:1998gx}.
Self-interactions of the $\phi$ field, as well as possible interactions
with other fields in the medium, fully justify the inclusion of
dissipation and noise terms such as done in the framework adopted here.

\begin{figure}[htb]
\vspace*{-0.1cm}
                 \insertplot{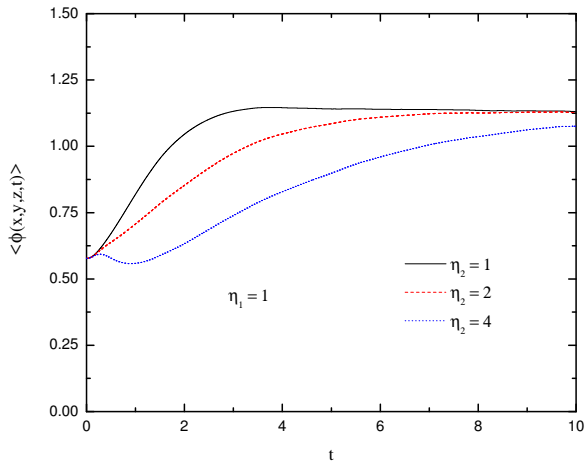}
\vspace*{-0.1cm}
\caption[]{Average value of the field $\phi$ as a function of time 
under the effect of additive and multiplicative noises 
($\eta_1=1$ and $\eta_2=1$, $2$, $4$)}
\label{multi-noise}
\end{figure}

In our numerical simulations we solve Eq.~(\ref{langevin}) on a
cubic space-like lattice. We use the leap frog method (see {\it e.g.}
Refs. \cite{Borrill:1996uq,Gagne:1999nh}) for the time evolution
and treat the Laplacian using a {}Fast {}Fourier Transform for the
spatial dependence~\cite{mimc}. To minimize lattice artifacts, we
use counterterms calculated in lattice perturbation in the
classical theory (tadpole plus setting-sun
diagrams)~\cite{Farakos:1994kx}:
\begin{equation} 
V_{CT} = \frac{1}{2}\left\{- 3 \times 0.252731 \frac{T}{a} + \frac{6
T^2}{(4\pi)^2} \left[\log \left(\frac{6}{\mu a } \right) + 0.009
\right] \right\} \phi^2 \; , 
\end{equation}
where $T$ is the temperature, $a$ is the lattice spacing and $\mu$
is an arbitrary renormalization scale. It should be noted that the 
addition of this counterterm is supposed to guarantee independence 
of the results on $a$ at {\em equilibrium} only. We checked the
lattice independence of the results \cite{FKR}, and adopt units such 
that $m =1$ in the Lagrangian of Eq.~(\ref{lagrangian}), {\it i.e.} 
all dimensional quantities are expressed in units of $m$. In our 
simulations, the lattice extent is fixed to $L=16$, and the time step 
is $\Delta t = 10^{-3}$; $T=1$, $\lambda =1$, and $\mu = 1$. 
The initial configuration is taken to
be a random distribution of the form $\phi(x,t) = \phi_{sp} +
0.001*rand(seed)$, where $\phi_{sp}=1/\sqrt{3}$ is the spinodal point
of the bare double-well potential, and $rand(seed)$ is a uniform 
random number distribution in the interval $[-1,1]$. At the end, 
results are averaged over several runs for different realizations
of the noises. 

In Figure~\ref{multi-noise} we present preliminary results for
the volume average of $\phi$ as a function of time for 
different values of the multiplicative dissipation parameter $\eta_2$. 
For simplicity, $\eta_1$ and $\eta_2$ are taken to be constants. More 
realistic parametrizations will appear soon \cite{FKR}. As $\eta_2$
increases, the effect is of delaying the equilibration of the
order parameter, as expected. 

%%%%%%%%%%%%%%%%%%%%%%%%%%%%%%%%%%%%%%%%%%%%%%%
\section{Non-Markovian corrections}

As discussed before, the structure of memory integrals and 
colored noise that appear in a realistic field-theoretic 
description of the dynamics of phase transitions is often rather 
complicated \cite{Arrizabalaga:2005tf,Berges:2005yt}. 
To develop and test new approaches, one 
can consider the much simpler case of  dissipation in quantum 
mechanics, where all the approximations and important scales 
are under control, and where one also finds inumerous 
applications \cite{weiss}.

Starting from the nonequilibrium evolution of a particle coupled 
linearly to a set of harmonic oscillators in the Caldeira-Leggett 
fashion \cite{weiss}, we study the effects of the 
non-local dissipation kernel, as well as of colored noise, 
that appear in the complete Langevin equation for the particle 
coordinate in space \cite{FKKP}. There, the memory kernel has 
its origin in the Feynman influence functional of the heat bath. 
To approach the kernel in a simpler, analytic way, we develop 
a systematic expansion in time derivatives whose convergence 
is regulated by increasing powers of the frequency cutoff in the 
distribution of oscillators, $\Omega$. Using this method, one can 
incorporate consistently corrections coming from the memory integral. 
The equation of motion thus obtained has the following form 
\cite{FKKP}:
\begin{equation}
M \ddot{Q}+ {V}'\left( Q \right) + 
\frac{2 \eta}{\pi} \sum^{\infty}_{n = 0} 
\frac{I_{n} \left( \Omega t \right)}{n!} 
~\frac{Q^{(n+1)} \left( t \right)}{\Omega^n} 
= \xi \left( t \right) \,, 
\label{eq2}
\end{equation}
which reduces to the traditional Langevin equation with white 
noise in the limit $\Omega t \rightarrow \infty$, since 
$I_{n} \left( \infty \right) \rightarrow \delta_{n0} \pi/2$ 
and the noise correlator tends to 
$2\eta T \delta \left( t-t' \right)$, consistently with the 
fluctuation-dissipation theorem. $I_{n}\left(\Lambda\right)$ 
are integral coefficients which can be expressed in terms of 
incomplete gamma functions. Inspection of (\ref{eq2}) shows 
that terms containing higher-order time derivatives of $Q$ 
are strongly suppressed by increasing powers of $1/\Omega$. 

%%%%%%%%%%%%%%%%%%%%%%%%%%%%%%%%%%%%%%%%%%%%%
\section{Final remarks}

The construction of a microscopic field-theoretic framework 
to study the role of noise and dissipation in a realistic, 
yet efficient, way is still not accomplished. We believe this 
task will require the development of systematic (controlled) 
analytic approximations to simplify memory kernels, as well as 
effective numerical methods to deal with arbitrary colored noise. 
In the case of the chiral transition in heavy ion collisions, 
effects brought about by the expansion of the plasma \cite{explosive} 
and by its finite size \cite{Fraga:2003mu} will also bring 
corrections to this picture. Some of these issues will be addressed 
in future publications \cite{FKR,FKKP}.

%%%%%%%%%%%%%%%%%%%%%%%%%%%%%%%%%%%%%%%%%%%%%
 
\section*{Acknowledgments}
My special thanks to T. Kodama, G. Krein, L. F. Palhares and 
R. O. Ramos for stimulating discussions and collaboration. 
The author also thanks the organizers of the 
QGP Thermalization Workshop for providing such a nice environment 
for the debate of ideas. The work of E.S.F. is partially supported by 
CAPES, CNPq, FAPERJ, and FUJB/UFRJ.
 
%\section*{Notes} (footnotes will be typeset in this way) 
%\begin{notes}
%\item[a]
%E-mail: author@host.domain.name
%\end{notes}

\vfill\eject
\end{document}